\documentclass[12pt]{iopart}
\usepackage{xcolor}
\usepackage{graphicx}% Include figure files
\usepackage{dcolumn}% Align table columns on decimal point
\usepackage{bm}% bold math
\usepackage{multirow}
\usepackage[numbers,sort&compress,square]{natbib}
\usepackage{acronym}
\usepackage{hyperref}% add hypertext capabilities
\usepackage{verbatim}
\usepackage{iopams} 
\usepackage[mathlines,running]{lineno}
\usepackage{ulem}
\graphicspath{{./final_figs/}}

\newcommand{\eqconst}{\stackrel{c}{=}}
\newcommand{\CIERA}{Center for Interdisciplinary Exploration and Research in Astrophysics (CIERA), Northwestern University, Evanston, IL 60201, USA}
\newcommand{\Princeton}{Department of Physics, Princeton University, Princeton, NJ 08544, USA}
\newcommand{\PrincetonRC}{Research Computing, Princeton University, Princeton, NJ 08544, USA}
\newcommand{\Austin}{Weinberg Institute, University of Texas at Austin, Austin, TX 78712, USA}
\newcommand{\UChicago}{Kavli Institute for Cosmological Physics, The University of Chicago, Chicago, IL 60637, USA}
\newcommand{\Utrecht}{Institute for Gravitational and Subatomic Physics (GRASP), Utrecht University, 3584 CC Utrecht, The Netherlands}
\newcommand{\Nikhef}{National Institute for Subatomic Physics (Nikhef), 1098 XG Amsterdam, The Netherlands}
\newcommand{\Portsmouth}{Institute of Cosmology \& Gravitation, University of Portsmouth, Portsmouth, PO1 3FX, United Kingdom}
\newcommand{\Glasgow}{Institute for Gravitational Research, School of Physics \& Astronomy, University of Glasgow, Glasgow, G12 8QQ, United Kingdom}
\newcommand{\ICREA}{Catalan Institution for Research and Advanced Studies (ICREA), E-08010 Barcelona, Spain}
\newcommand{\IFAE}{Institut de F\'{ı}sica d’Altes Energies (IFAE), The Barcelona Institute of Science and Technology, UAB Campus, E-08193 Barcelona, Spain}
\newcommand{\GSSI}{Gran Sasso Science Institute (GSSI), I-67100 L'Aquila, Italy}
\newcommand{\INFN}{INFN, Laboratori Nazionali del Gran Sasso, I-67100 Assergi, Italy}
\newcommand{\CIT}{Department of Physics, California Institute of Technology, Pasadena, California 91125, USA}
\newcommand{\CITA}{Canadian Institute for Theoretical Astrophysics, University of Toronto, 60 St. George Street, Toronto, ON M5S 3H8, Canada}
\newcommand{\CCA}{Center for Computational Astrophysics, Flatiron Institute, New York NY 10010, USA}
\newcommand{\StonyBrook}{Department of Physics and Astronomy, Stony Brook University, Stony Brook NY 11794, USA}
\usepackage{iopams}
\begin{document}

% \title[Windows in GW Inference]{Biased and unbiased likelihood functions for gravitational-wave parameter estimation with windowed data}
% \title[Windows in GW Inference]{Inference with finite time series II: Biased and unbiased likelihood functions for gravitational-wave parameter estimation with windowed data}
\title[Inference with finite time series II]{Inference with finite time series II: the window strikes back}

% \author{}

% \email{aditya@utoronto.ca}

% AV: \orcidlink{0000-0002-4103-0666}

% \affiliation{}

\author{
    Colm Talbot$^{1,2,3}$,
    Sylvia Biscoveanu$^{1,4}$\footnote{NASA Einstein Fellow},
    Aaron Zimmerman$^{5}$,
    Tomasz Baka$^{6,7}$,
    Will M.~Farr$^{8,9}$,
    Jacob Golomb$^{10}$,
    Charlie Hoy$^{11}$,
    Andrew Lundgren$^{12,13}$,
    Jacopo Tissino$^{14,15}$,
    John Veitch$^{16}$,
    Aditya Vijaykumar$^{17}$
    Michael J. Williams$^{11}$,
}
\address{$^{1}$\Princeton}
\address{$^{2}$\PrincetonRC}
\address{$^{3}$\UChicago}
\address{$^{4}$\CIERA}
\address{$^{5}$\Austin}
\address{$^{6}$\Utrecht}
\address{$^{7}$\Nikhef}
\address{$^{8}$\StonyBrook}
\address{$^{9}$\CCA}
\address{$^{10}$\CIT}
\address{$^{11}$\Portsmouth}
\address{$^{12}$\ICREA}
\address{$^{13}$\IFAE}
\address{$^{14}$\GSSI}
\address{$^{15}$\INFN}
\address{$^{16}$\Glasgow}
\address{$^{17}$\CITA}
\ead{colm.talbot@princeton.edu}
\vspace{10pt}
\begin{indented}
\item[]\today
\end{indented}

\begin{abstract}
Smooth window functions are often applied to strain data when inferring the parameters describing the astrophysical sources of gravitational-wave transients.
Within the LIGO-Virgo-KAGRA collaboration, it is conventional to include a term to account for power loss due to this window in the likelihood function.
We show that the inclusion of this factor leads to biased inference.
The simplest solution to this, omitting the factor, leads to unbiased posteriors and Bayes factor estimates provided the window does not suppress the signal for signal-to-noise ratios $\lesssim O(100)$, but unreliable estimates of the absolute likelihood.
Instead, we propose a multi-stage method that yields consistent estimates for the absolute likelihood in addition to unbiased posterior distributions and Bayes factors for signal-to-noise ratios $\lesssim O(1000)$.
Additionally, we demonstrate that the commonly held wisdom that using rectangular windows necessarily leads to biased inference is incorrect.
\end{abstract}

\vspace{2pc}
\noindent{\it Keywords}: gravitational waves, Bayesian inference, time series analysis
%\submitto{\CQG}
%\maketitle
\acrodef{O4}[O4]{fourth observing run}
\acrodef{BBH}[BBH]{binary black hole}
\acrodef{GW}[GW]{gravitational-wave}
\acrodef{IID}[IID]{independently and identically distributed}
\acrodef{LVK}[LVK]{LIGO-Virgo-KAGRA}
\acrodef{PN}[PN]{post-Newtonian}
\acrodef{SOR}[SOR]{spin-orbit resonance}
\acrodef{RMR}[RMR]{reversed mass ratio}
\acrodef{SMR}[SMR]{standard mass ratio}
\acrodef{SNR}[SNR]{signal-to-noise ratio}
\acrodef{VM}[VM]{von Mises}
\acrodef{KDE}[KDE]{Kernel Denisty Estimate}
\acrodef{MC}[MC]{Monte Carlo}
\acrodef{MCMC}[MCMC]{Markov Chain Monte Carlo}
\acrodef{CDF}[CDF]{cumulative distribution function}
\acrodef{PPD}[PPD]{posterior predictive distribution}
\acrodef{PSD}[PSD]{power spectral density}
\acrodef{ACF}[ACF]{autocorrelation function}
\acrodef{PP}{probability-probability}

\section{Introduction} 
\label{sec:intro}
The characterization of gravitational-wave sources via Bayesian inference is key to understanding their properties (e.g.,~\cite{KAGRA:2021vkt}; see, e.g., \cite{Christensen:2022bxb, Chatziioannou:2024hju, Roulet:2024cvl} for recent reviews).
Accurate and precise constraints on the parameters of merging compact binaries are needed to facilitate tests of general relativity~\cite{LIGOScientific:2021sio, Gupta:2024gun}, searches for gravitational-wave lensing signatures~\cite{LIGOScientific:2023bwz}, multimessenger follow-up of gravitational-wave sources~\cite{LIGOScientific:2017ync}, independent measurements of cosmological parameters~\cite{LIGOScientific:2017adf, LIGOScientific:2021aug}, and probes of the astrophysics of the population of compact-object mergers as a whole~\cite{KAGRA:2021duu}. 

In the absence of astrophysical or terrestrial transients, gravitational-wave strain data from ground-based interferometers like the \ac{LVK} detectors~\cite{LIGOScientific:2014pky, VIRGO:2014yos, KAGRA:2020tym} are typically Gaussian and stationary on timescales of seconds--minutes. If the data are also periodic, the noise covariance matrix is diagonal in the frequency domain.
While the time- and frequency-domain data representations are equivalent and related by the Fourier transform, \ac{GW} analyses are typically performed in the frequency domain because this additional simplicity reduces the computational burden of computing the likelihood function. 
However, Bayesian inference analyses of \ac{GW} transients are restricted to a finite stretch of data containing the signal in the sensitive frequency band of the instruments (see \cite{LIGOScientific:2019hgc} for an overview of \ac{LVK} data analysis methods). 
As the data in this case are a subset of a much longer data stream, they are not periodic as required for the Fourier basis to diagonalize the noise covariance matrix.
Therefore, the noise covariance matrix is not perfectly diagonal in the frequency domain~\citep{Talbot:2021igi}.
To mitigate this effect, a smooth window function can be applied that gradually sends the value of the data to zero at the edges of the segment~\cite{tukey1967spectrum, harris1978use}.
This window suppresses the total noise power in the segment, and it is common to renormalize the data to preserve the total noise power, which is equivalent to modifying the likelihood function to account for the power loss~\cite[e.g.,][]{Veitch:2014wba, Cornish:2014kda, Ashton:2018jfp, Dax:2021tsq}.

In this paper, we demonstrate that this correction to the likelihood should not be included and describe the extent to which results are biased with this error.
In Section~\ref{sec:formalism}, we describe the formalism used for \ac{GW} data conditioning and parameter estimation analyses and the faulty motivation for correcting the likelihood to account for the power loss due to windowing. We justify the use of the form of the likelihood without this correction via tests on \ac{GW} data in Section~\ref{sec:demonstration}, present an alternative analysis method in Section~\ref{sec:solutions}, and conclude in Section~\ref{sec:conclusion}.

\section{Formalism}
\label{sec:formalism}
\ac{GW} strain data can be represented in the time domain as a sum of contributions from detector noise, $n(t)$ and the response of the detectors to astrophysical signals, $h(t)$,
\begin{equation}
    d_{i} = n_{i} + h_{i}(\theta), \label{eq:strain}
\end{equation}
where the continuous timeseries are discretely sampled such that $x_i = x(t_i)$.
For a quasi-circular compact-object binary, $\theta$ includes parameters intrinsic to the compact objects, e.g., mass, extrinsic parameters that determine the relative position and orientation of the source with respect to the observer, and calibration parameters describing the response of the detector to the signal. For a data segment with duration $T$ sampled at a frequency $f_{s}$ consisting of $N=Tf_{s}$ independent measurements, the data can be equivalently represented in the frequency domain using the discrete Fourier transform,
\begin{equation}
    \tilde{d}_{k} = \frac{1}{f_{s}}\sum_{j=0}^{N-1}d_{j}e^{-2\pi ijk/N}, \quad k = 0, 1, ..., N/2-1.
\end{equation}
Here, the frequency series are discretely sampled at frequencies $\tilde{x}_{k} = \tilde{x}(k / T)$.
Under the assumption that the noise is stationary and Gaussian, the likelihood of obtaining a given realization is described in either the time or frequency domain,
\begin{eqnarray}
    \log \mathcal{L}(n) &\eqconst -\frac{1}{2} \sum_{i,j=0}^{N} n_{i} C^{-1}_{ij} n_{j},\\
    \log \mathcal{L}(\tilde{n}) &\eqconst -\frac{1}{2} \sum_{i,k=0}^{N/2-1} \tilde{n}_{i}^{*} \tilde{C}^{-1}_{ik} \tilde{n}_{k}, \label{eq:freq_domain_likelihood}
\end{eqnarray}
where the time-domain covariance matrix for a stationary random process is a symmetric Toeplitz matrix given by the autocovariance function,
\begin{equation}
    C_{ij} = C(\tau_{i-j}) = \mathrm{cov}[n(t_{i}), n(t_{i}-\tau_{i - j})],
\end{equation}
for $\tau_{i - j} = t_{i} - t_{j}$.
Here and throughout, the symbol $\eqconst$ means that two quantities are equal up to an additive constant.

\begin{figure*}
\begin{center}
\includegraphics[width=\linewidth]{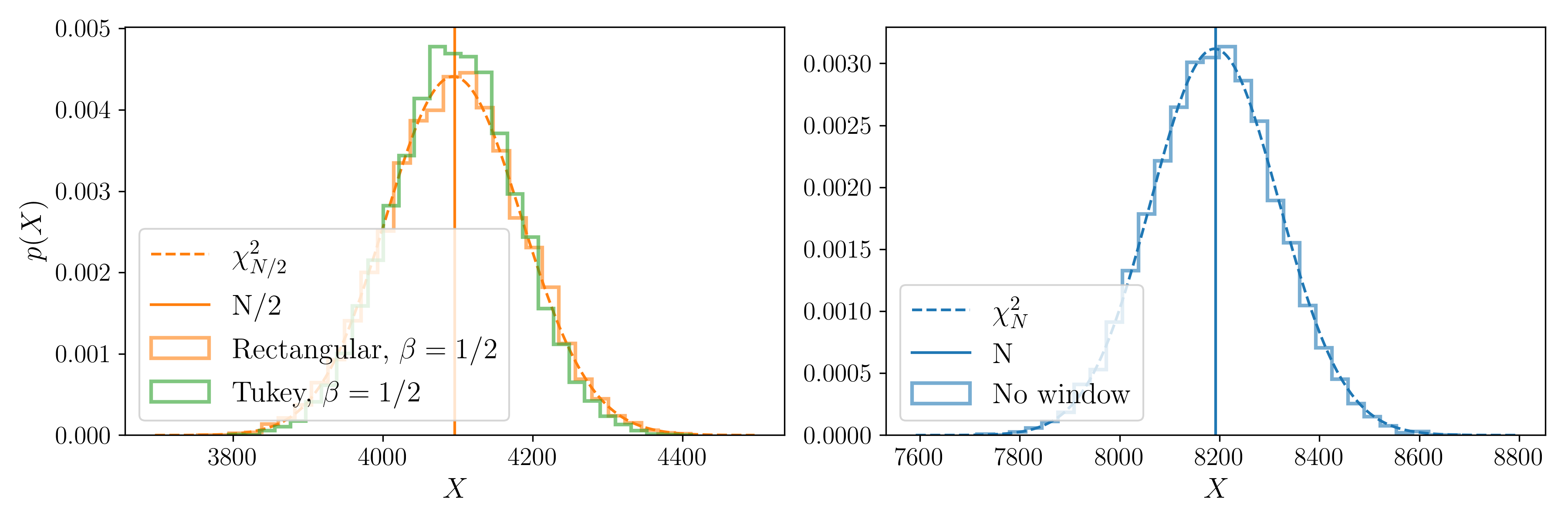}
\caption{Empirically-calculated distributions of $X$ as defined in \ref{eq:def_chisq} for 10,000 realizations of $4~\mathrm{s}$ of white, Gaussian noise sampled at $2048~\mathrm{Hz}$ ($N=8192$) with no window (blue, right), a rectangular window with $\beta=1/2$ (orange, left), and a Tukey window with $\beta=1/2$ (green, left) compared to the expected $\chi^{2}$ distributions (dashed lines).}\label{fig:chisq}
\end{center}
\end{figure*}

If the data are additionally periodic with period $T$, the frequency-domain noise covariance matrix is diagonal, given by the one-sided \ac{PSD}, $S(f)$~\cite{unser1984approximation}, 
\begin{equation}
    \tilde{S}_k = \frac{4}{T} \delta_{ik} \tilde{C}_{ik}.
    \label{eq:freq_cov}
\end{equation}
Another special case where the noise covariance matrix is diagonal is when the noise is white, i.e., $S_k = S$.
The frequency-domain likelihood in \ref{eq:freq_domain_likelihood} is the Whittle likelihood~\cite{whittle1957curve, whittle1962gaussian}, which for an ensemble of data points $\tilde{n} = (\tilde{n}_{0}, \tilde{n}_{1}, ... \tilde{n}_{N/2-1})$ can be written as
\begin{equation}
    \log \mathcal{L}(\tilde{n}) = \sum_{k=0}^{N/2-1}\log{\left[\frac{2}{\pi T S_{k}}\right]} -\frac{2}{T}\frac{\mathfrak{R}(\tilde{n}_{k})^{2} + \mathfrak{I}(\tilde{n}_{k})^{2}}{S_{k}} \label{eq:whittle}
\end{equation}
Assuming that the real and imaginary components of each $\tilde{n}_{k}$ are independent, the rightmost term in \ref{eq:whittle}
is $\chi^{2}$-distributed with $N$ degrees of freedom,
\begin{eqnarray}
    X &= \sum_{k=0}^{N/2-1}\frac{2(\mathfrak{R}(\tilde{n}_{k})^{2} + \mathfrak{I}(\tilde{n}_{k})^{2})}{TS_{k}} \label{eq:def_chisq}\\
    X &\sim \chi^{2}_{N}.
    \label{eq:chisq}
\end{eqnarray}
The right panel of Figure~\ref{fig:chisq} shows a comparison between the $\chi^{2}_{N}$ distribution and the distribution obtained by calculating $X$ for 10,000 realizations of white Gaussian noise.

Analyses of transient \ac{GW} signals are typically restricted to a finite stretch of data containing the signal while in the sensitive frequency band of the interferometers rounded to a power of two in seconds.
Additionally, the sum in ~\ref{eq:whittle} runs over frequencies between $\sim 20 - 2000$ Hz as the data are known not to obey the expected statistics and the signal generally has a much smaller amplitude than the noise outside this range.
This data selection can be achieved while minimizing the impact of discontinuities due to non-periodicity of the data by applying a smooth window function that sets the data to zero outside the time range of interest~\cite{harris1978use}.
The application of the window function suppresses the power in the data by a factor equal to the power in the window function,
\begin{equation}
    \beta = \frac{1}{T}\int dt \, w^2(t),
\end{equation}
and introduces artificial non-stationarity, changing the noise statistics.
As an example, consider a unitary rectangular window applied to uncorrelated data with $N$ independent time samples that zeros the first and last quarter of the data samples such that $\beta = 1/2$.
After windowing, $X \sim \chi^{2}_{N/2}$ for this stretch of data, as there are only $N/2$ independent data points in the time domain (orange histogram and distribution in Figure~\ref{fig:chisq}).
Generically, the whitened frequency-domain strain will follow a $\chi^{2}$ distribution with $N\beta$ degrees of freedom, $X \sim \chi^{2}_{N\beta}$ if a rectangular window is applied to periodic data.
However, non-rectangular windows introduce correlations between frequency bins, such that the frequency-domain covariance matrix in \ref{eq:freq_cov} is no longer diagonal~\cite{Talbot:2021igi}.
Thus, the whitened frequency-domain data are no longer $\chi^{2}$-distributed, as the assumption of independent random variables underlying \ref{eq:chisq} no longer holds (see the green histogram in Figure~\ref{fig:chisq} corresponding to a Tukey window).

\begin{figure*}
\begin{center}
\includegraphics[width=\linewidth]{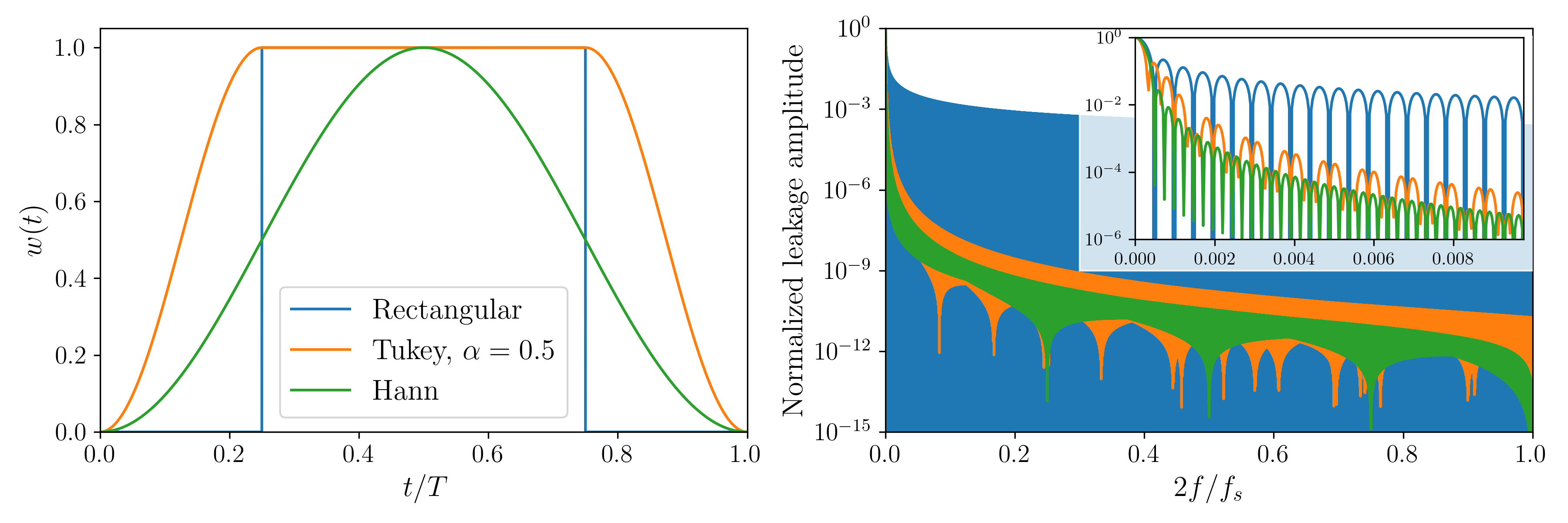}
\caption{ \textit{Left:} Tukey windows with different $\alpha$ values in the time domain: $\alpha=0$ (rectangular) in blue, $\alpha=0.5$ in orange, and $\alpha=1$ (Hann) in green. \textit{Right:} Normalized leakage amplitude, $|\tilde{w}(f)|/|\tilde{w}(0)|$, for the three windows on the left over the entire frequency domain and zoomed in to $f = 20/f_{s}$ in the inset.}\label{fig:windows}
\end{center}
\end{figure*}

In \ac{GW} astronomy, the Tukey window~\cite{tukey1967spectrum} is commonly used:
\begin{equation}
    w(t) = \cases{
        \frac{1}{2} \left[ 1 - \cos\left( \frac{2\pi t}{\alpha T} \right) \right], & $0 \leq t \leq \frac{\alpha T}{2}$\\
        1, & $\frac{\alpha T}{2} < t \leq T \left(1 - \frac{\alpha}{2} \right)$\\
        \frac{1}{2} \left[ 1 + \cos\left( \frac{2\pi}{\alpha T}\left(t - T\left(1 - \frac{\alpha}{2} \right)\right) \right) \right], & $T \left(1 - \frac{\alpha}{2} \right) < t \leq T$
    }.
\end{equation}
The free parameter $\alpha$ is the fraction of the data segment where $w(t)<1$; in the $\alpha=0$ limit, the Tukey window becomes a rectangular window, and for $\alpha=1$ it becomes a Hann window. 
For the Tukey window, $\alpha$ is related to the fractional power loss parameter,
\begin{equation}
    \beta = 1 - \frac{5\alpha}{8}.
\end{equation}
The resulting windowed frequency series is the convolution of the Fourier transforms of the window function and the time-domain data,
\begin{eqnarray}
    d_{w,j} &= w_{j}d_{j}, \\
   \tilde{d}_{w,k} &= \frac{1}{f_{s}}\sum_{j=0}^{N-1}w_{j} d_{j} e^{-2\pi ijk/N} = (\tilde{d} * \tilde{w})_{k}, \quad 0 \leq k \leq N/2-1.
\end{eqnarray}
In Figure~\ref{fig:windows}, we show three window functions in the time (left) and frequency (right) domains.
Specifically, we show a rectangular window that truncates the data to the central $50\%$ of the segment (blue), a Tukey window with $\alpha=0.5$ (orange), and a Hann window (green).
In the frequency domain, all three windows exhibit a clear pattern of peaks and troughs separated by the duration of the window.
The Fourier transform of a rectangular window that is non-zero for a duration $T_w$ is the sinc function, which is exactly zero at all multiples of $1 / T_w$.
The troughs for the Tukey and Hann windows are spaced twice as narrowly, as these are defined over twice the duration as the rectangular window in this example.
Since the rectangular window goes exactly to zero at these points, if the data are periodic over period $T_w$, a rectangular window covering that duration will induce no correlation between frequencies.

However, real \ac{GW} strain data are not periodic and exhibit sharp spectral features, or ``lines,'' in the \ac{PSD}~\cite{LSC:2018vzm, LIGO:2024kkz}. 
Both of these features of the data contribute to spectral leakage: 1) The compactness of the analysis data in the time domain imposed by the window translates into infinite support in the frequency domain, and 2) Unless the lines occur at frequencies that are integer multiples of $1/T$, the discontinuities introduced when the data are windowed mean that some of the power in these lines leaks into neighboring frequencies.
In this case, it is common to choose a window function that minimizes the spectral leakage.
We see that the spectral leakage decays much more rapidly for smoothly varying windows than for the rectangular window (see Appendix D of \cite{Romano:2016dpx} for a more comprehensive discussion of different window functions).

There is a trade-off between the suppression of spectral leakage and power loss in the data for each choice of window.
To account for this power loss, many flagship \ac{GW} parameter estimation software packages~\cite{Veitch:2014wba, Cornish:2014kda, Ashton:2018jfp, Dax:2021tsq} modify the likelihood in \ref{eq:whittle} to include a factor of $\beta$~\cite{talbot2020thesis},
\begin{equation}\label{eq:biased-likelihood}
    \ln {\cal L}_{\beta}(\tilde{d}|\theta) \eqconst \sum_{k=0}^{N/2-1} -\frac{2|\tilde{d}_k - h_k(\theta)|^{2}}{\beta T S_k},
\end{equation}
\begin{equation}\label{eq:biased-likelihood}
    \ln {\cal L}(\tilde{d}|\theta) \eqconst \sum_{k=0}^{N/2-1} -\frac{2|\tilde{d}_k - h_k(\theta)|^{2}}{T S_k},
\end{equation}
where we have used $\tilde{n}_{k} = \tilde{d}_{k} - \tilde{h}_{k}(\theta)$ implied by \ref{eq:strain}\footnote{This same window factor is required to correct the \ac{PSD} when it is estimated from data which has been windowed.
It is customary in the \ac{GW} parameter estimation literature to apply the same window to the analyzed segment of data and data used to estimate the \ac{PSD} to ensure that spectral leakage is consistent between the two cases.}.
The posterior distribution over the signal model parameters $\theta$ is then obtained using Bayes' theorem,
\begin{equation}
    p(\theta | \tilde{d}) = \frac{\mathcal{L}_{\beta}(\tilde{d} | \theta)\pi(\theta)}{\mathcal{Z}},
\end{equation}
where $\pi(\theta)$ is the prior on the signal model parameters and $\mathcal{Z}$ is the evidence, or marginalized likelihood.

To ensure the theoretical signal model agrees with the true signal, the analysed data segment is chosen so that the window has minimal impact on the signal.
For a compact-object binary, this is typically achieved by placing the point estimate for the moment of the merger from the matched-filter search pipelines~\cite{Allen:2005fk, Messick:2016aqy, Adams:2015ulm, Usman:2015kfa} $2~\mathrm{s}$ from the end of the analysis segment and using a Tukey window with roll-off $r = T\alpha/2\leq 1~\mathrm{s}$.
Since the statistics of the noise \textit{in the part of the analysis segment that contains the signal} are not affected by the window, the $\beta$ factor should not be included in the likelihood in~\ref{eq:biased-likelihood}. 

To show this mathematically, we make explicit that the astrophysical contribution to \ref{eq:strain} in the sensitive frequency band, $h_{j}(\theta)$, is nonzero only for part of the segment, represented by $j \in \mathcal S$.
Then the log likelihood for the data can be written as
\begin{eqnarray}
    \ln \mathcal L(d | \theta) &\eqconst -\frac 12 \sum_{ij} [d_i - h_i(\theta) ] C^{-1}_{ij} [d_j - h_j(\theta) ],\\
    &= -\frac 12 \sum_{ij} d_{i} C^{-1}_{ij}d_{j} + \sum_{i \in \mathcal{S},j} h_{i}(\theta)C^{-1}_{ij}d_{j} - \frac{1}{2} \sum_{i,j \in \mathcal{S}} h_{i}(\theta)C^{-1}_{ij}h_{j}(\theta)
    .
\label{eq:likelihood_segmented}
\end{eqnarray}
From the perspective of the posterior, $\theta$-independent parts of the log likelihood are (data-dependent) multiplicative constants. In other words, if we already know the noise properties by knowing $C_{ij}$, our inference on $\theta$ is neither improved nor degraded by including draws from the noise in our dataset.

Now we consider the effect of windowing the data with a window that does not affect the part of the segment including the signal model, $w_j = 1\ \forall\ j \in \mathcal{S}$.
By construction, the window does not change the signal model, i.e.,
\begin{equation}
    h_j = w_j h_j.
\end{equation}
The data and covariance matrix are impacted by the window function
\begin{eqnarray}
    d_{w,j} = w_j d_j, &\quad \Rightarrow \quad d_w = D_w d \\
    C_{w,ij} = w_i w_j C_{ij}, &\quad \Rightarrow \quad C_w = D_w C D_w \,,
\end{eqnarray} 
where $D_w$ is a diagonal matrix whose entries are $D_{w,jj} = w_j$.
Note that with a multiplicative window, the Gaussian noise remains Gaussian.
Then the inverse noise covariance matrix can be expressed as
\begin{equation}
C_w^{-1} = D_w^{-1} C^{-1} D_w^{-1} \quad \Rightarrow \quad C^{-1}_{w,ij} = \frac{C^{-1}_{ij}}{w_i w_j} \,.
\end{equation}
Putting it all together, the likelihood of the windowed time-domain data is
\begin{eqnarray}
\ln \mathcal L_w(d_w | \theta)  &\eqconst -\frac 12 \sum_{ij} (d_{w,i} - h_i) C^{-1}_{w,ij} (d_{w,j} - h_j) \\
& = -\frac 12 \sum_{ij}  d_{w,i} C^{-1}_{w,ij} d_{w,j}
+ \sum_{i \in \mathcal S, j} h_i(\theta) C^{-1}_{ij} \frac{d_{w,j}}{w_j} \nonumber \\
& \qquad \qquad - \frac 12 \sum_{i,j \in \mathcal S} h_i(\theta) C^{-1}_{ij} h_j(\theta).
\label{eq:windowed_likelihood}
\end{eqnarray}
Since $d_{j} = d_{w,j}/w_{j}$, this is equivalent to the likelihood for the unwindowed data in \ref{eq:likelihood_segmented} up to the $\theta$-independent factors.

This result means that using the Whittle likelihood in the frequency domain without the $\beta$ factor is equivalent to using the likelihood $\mathcal{L}$ in \ref{eq:likelihood_segmented} but with the direct substitution $d_{j} \rightarrow d_{w,j}$. 
This is not equivalent to the correct likelihood $\mathcal{L}_w$ in \ref{eq:windowed_likelihood} since the non-diagonal inverse covariance can create cross terms where $w_j \neq 1$ but $j \in \mathcal S$,
\begin{eqnarray}
\sum_{i \in \mathcal S, j} h_i(\theta) C^{-1}_{ij} \frac{d_{w,j}}{w_j} =  \sum_{i \in \mathcal S, j} h_i(\theta) C^{-1}_{ij} d_j \neq \sum_{i \in \mathcal S, j} h_i(\theta) C^{-1}_{ij} d_{w,j} \,.
\end{eqnarray}
The unwindowed likelihood applied to the windowed data can thus be related to the correct windowed likelihood by
\begin{eqnarray}
\ln \mathcal L(d_w |\theta) &\eqconst -\frac 12 \sum_{ij} (d_{w,i} - h_i(\theta)) C^{-1}_{ij} (d_{w,j} - h_j(\theta))\\
&\eqconst \ln \mathcal L_w(d_w | \theta) - \sum_{i \in \mathcal S, j} h_i(\theta) C^{-1}_{ij} (d_j - d_{w,j}).
\end{eqnarray}
If the waveform model is sufficiently short and the window has a large range of times where it is unity, then for the $\theta$-dependent correction factor to be significant, the inverse covariance must couple the signal model to stretches of data that are far away in time.
Alternatively, for the approximation to be an issue, the whitened model (which spreads beyond $i \in \mathcal S$) must overlap with the parts of the whitened data affected by the window. 
To see this more clearly, we define the whitening filter $\mathcal W$ as the square root of $C^{-1}$, so $\mathcal W^\top \mathcal W = C^{-1}$. 
Then the problematic term is
\begin{eqnarray}
    \delta \ln {\cal L}(\theta) = \sum_{i \in \mathcal S, j} h_i(\theta) C^{-1}_{ij} (d_j - d_{w,j}) 
    &= 
    (\mathcal W h(\theta))^\top \mathcal W (1- D_{w})d .
\label{eq:correction_term}
\end{eqnarray}
Thus, whether this term is negligible or not depends on if the whitened signal model $\mathcal W h$ is spread out enough in time that it intersects with the portion of whitened data affected by the non-unity component of the window, i.e., $\mathcal W (1 - D_w) d \neq 0$.
If the whitened signal model and whitened, windowed data have minimal overlap, the frequency-domain Whittle likelihood with the windowed data $\tilde d_w$ and without the $\beta$ factor is approximately equivalent to the strictly correct likelihood $\mathcal{L}_w$ up to a $\theta$-independent factor, but is computationally cheaper to evaluate.

In general, inference is unbiased when systematic errors in log-likelihood estimates vary by less than $O(1)$ over the posterior support~\cite{Leslie:2021ssu, Talbot:2023pex}.
In our case, we can quantify this by calculating the standard deviation of $\delta \ln {\cal L}$ over the posterior support as a proxy for the variation in $\delta \ln {\cal L}$.
It is also informative to consider the dependence of $\delta \ln {\cal L}$ on the matched filter \ac{SNR}
\begin{equation}
    \rho(h | d) = \frac{\sum_{ij} d_i C^{-1}_{ij} h_j}{\sqrt{\sum_{ij} h_i C^{-1}_{ij} h_j}}.
\end{equation}
We can see that $\delta \ln {\cal L}\propto \rho$ and, equivalently, $\sigma_{\delta} \propto \rho$.
Our threshold on the \ac{SNR} at which our inference is likely biased is
\begin{eqnarray}
    \sigma_{\delta} &= \hat{\sigma}_{\delta} \frac{\rho_{\rm bias}}{\hat{\rho}} \approx 1 \\
    \rho_{\rm bias} &\approx \frac{\hat{\rho}}{\hat{\sigma}_{\delta}}.
\end{eqnarray}
Here we have defined the \ac{SNR} and typical bias at a reference \ac{SNR} denoted by $\hat{\rho}$ and $\hat{\sigma}_{\delta}$ respectively.

\begin{figure*}
\begin{center}
\includegraphics[width=\linewidth]{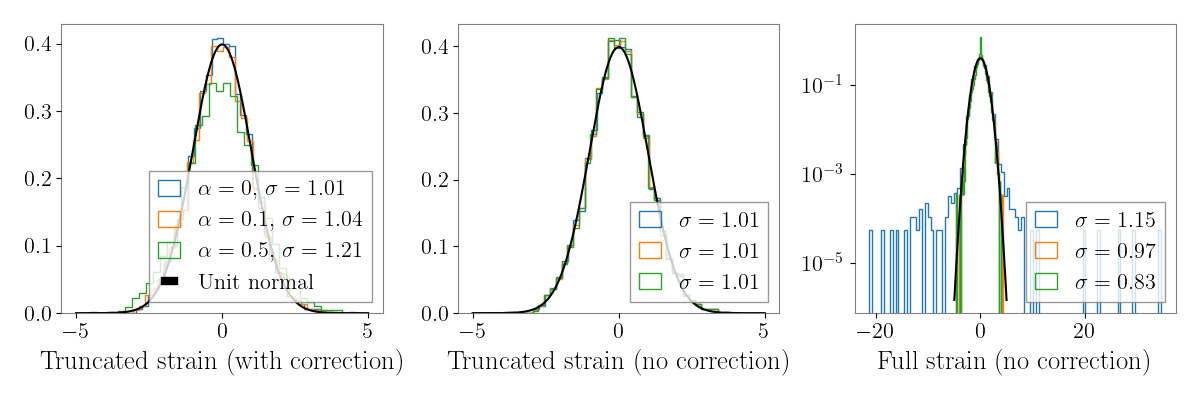}
\caption{Distribution of time-domain data from a $16~\mathrm{s}$ segment sampled at $2048~\mathrm{Hz}$ around the time of GW150914~\cite{LIGOScientific:2016aoc} in the LIGO Hanford detector after the data has been Tukey windowed with various $\alpha$ parameters including (excluding) the $\beta$ correction, Fourier transformed into the frequency domain, whitened, and inverse Fourier transformed back to the time domain in the left (middle) panel.
The legend shows the corresponding standard deviation of the samples $\sigma$.
The rightmost panel includes the times affected by the window, while the left and middle panels exclude those times. 
The data only follow a unit Gaussian in the middle panel, without the $\beta$ correction and limited to the times unaffected by the window.}\label{fig:whitened_norm_comp}
\end{center}
\end{figure*}

To demonstrate the impact of different window functions graphically, we consider a stretch of data that has been Tukey windowed in the time domain, Fourier transformed into the frequency domain, whitened, and then inverse Fourier transformed back into the time domain.
In the left panel of Figure~\ref{fig:whitened_norm_comp}, we show the distribution of the whitened time-domain data restricted to the part of the segment not affected by the window after dividing by $\sqrt{\beta}$ to account for the power loss.
We see that the distribution deviates increasingly from a unit Gaussian distribution as $\alpha$ increases.
Meanwhile, the analogous distributions of data whitened in the frequency domain without including the $\beta$ factor shown in the middle panel are all unit Gaussian-distributed.
However, the noise statistics of the \textit{segment as a whole} are affected by the window.
In the right-hand panel, we show the distribution of the whitened noise in the full segment.
For all windows, the distribution is not Gaussian.
As argued mathematically above, this means that the likelihood in \ref{eq:whittle} is formally incorrect when used with windowed data, even when the $\beta$ factor is omitted.
Below, we demonstrate that the correction term in \ref{eq:correction_term} is not significant for most applications in \ac{GW} parameter estimation---meaning that the form of the Whittle likelihood with $\beta$ omitted is sufficient when only the signal-to-noise likelihood \textit{ratio} is of interest---and discuss possible solutions to formally correct the likelihood.

\section{Demonstration for gravitational-wave parameter estimation}
\label{sec:demonstration}
In this section, we perform two tests that demonstrate the bias in \ref{eq:biased-likelihood} and that this bias can be mitigated by simply removing the $\beta$ factor.
The first is a \ac{PP} test, widely used in gravitational-wave astronomy to demonstrate unbiased inference, which shows the \ac{CDF} of the credible interval at which the true value of a parameter is recovered for an ensemble of simulated signals as a function of the credible interval.
For unbiased inference, the \ac{CDF} should follow a uniform distribution, which can be quantified with a p-value.
We simulate 5000 $4~\mathrm{s}$ segments of data consisting of a \ac{BBH} signal added to Gaussian noise colored by the \acp{PSD} expected for the LIGO Hanford, Livingston, and Virgo detectors during the \ac{O4} sampled at $512~\mathrm{Hz}$ that is then Tukey-windowed with $\alpha = 0.5$ ($\beta=0.6875$).
For simplicity, we vary only one parameter that has a simple effect on the signal in our test, the luminosity distance.
The \ac{BBH} systems are non-spinning and have fixed masses ($m_{1}=66~M_{\odot},\ m_{2}=59~M_{\odot}$), with distances distributed uniformly between $10\mbox{--}1000~\mathrm{Mpc}$.
The other extrinsic parameters are fixed to arbitrary values.
We perform the test with the same simulated data with and without the $\beta$ likelihood correction.
The likelihood is evaluated on a grid analytically in one dimension.

\begin{figure*}
\begin{center}
\includegraphics[width=0.49\linewidth]{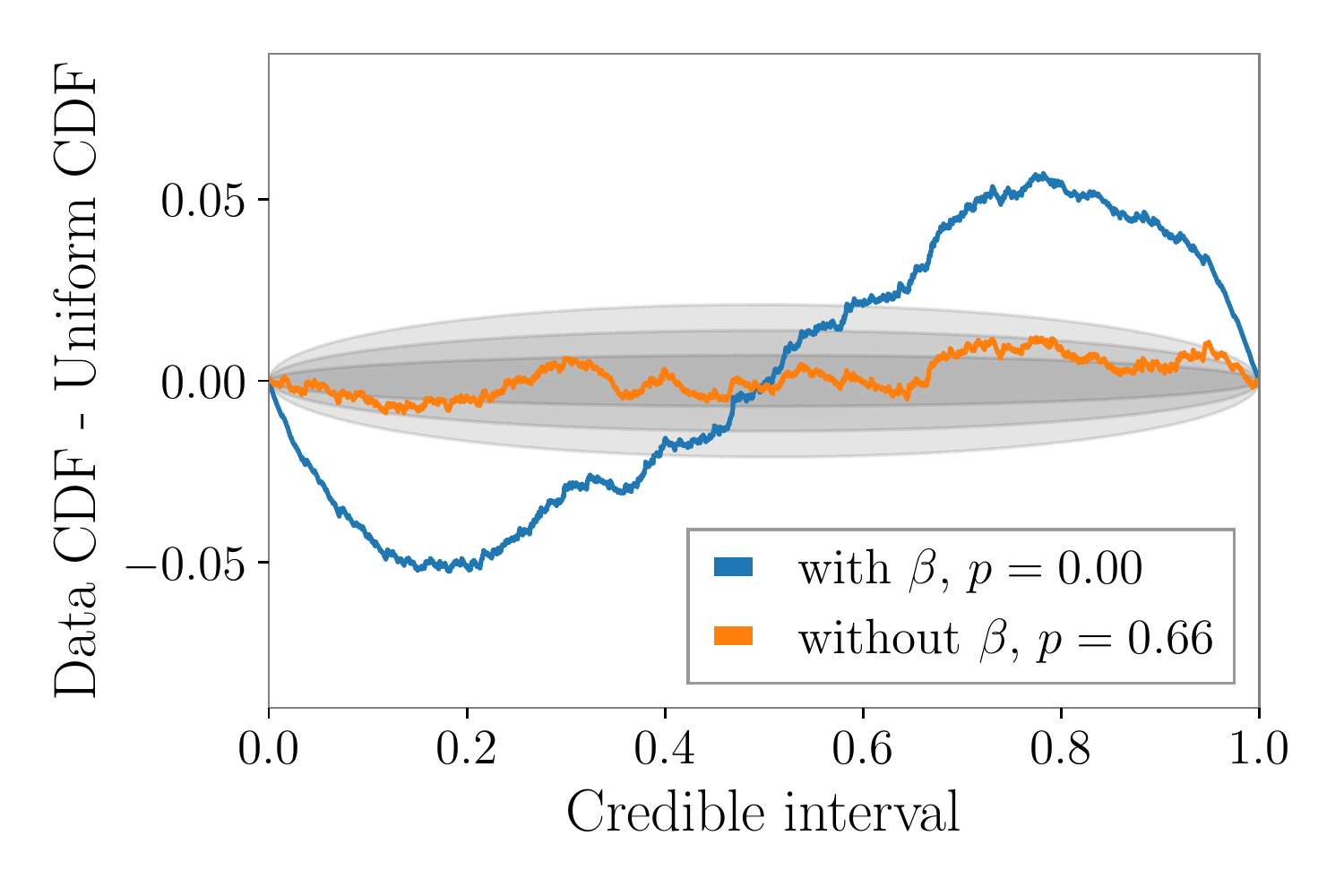}
\includegraphics[width=0.49\linewidth]{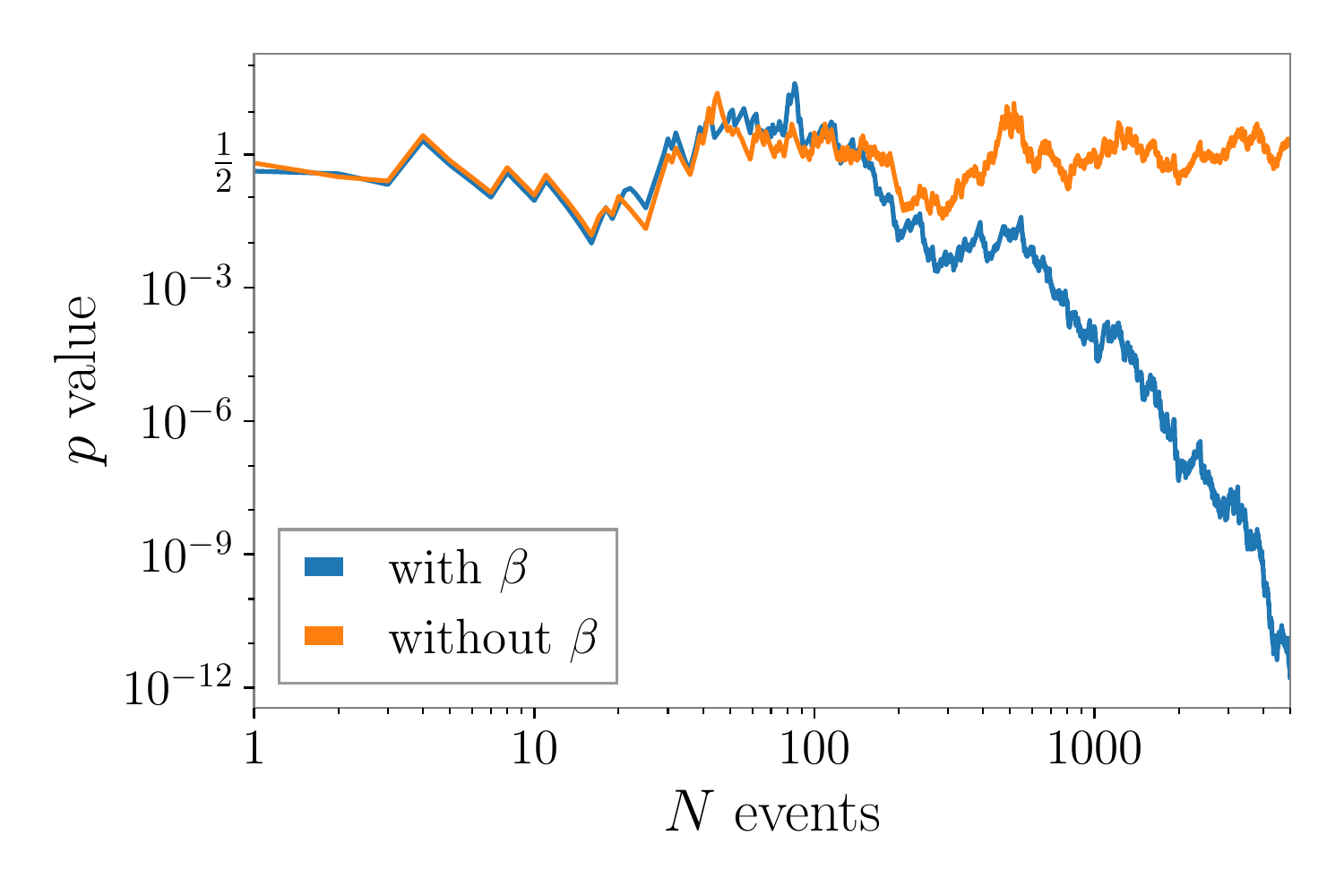}
\caption{
\textit{Left:} \ac{PP} plot for luminosity distance recovery for \ac{BBH} injections into Gaussian noise with (blue) and without (orange) the $\beta$ correction factor.
The grey shaded regions show the 1-, 2-, and 3-$C$ uncertainty regions.
After removing the correction, the $p$ value indicates that the results are entirely consistent with the null hypothesis.
\textit{Right:} the evolution of the $p$ value with the number of events for the two cases.
We see that when applying the $\beta$ likelihood correction, the test begins to fail after a few hundred events are included.
}\label{fig:pp-test}
\end{center}
\end{figure*}

The results are shown in Figure~\ref{fig:pp-test}.
The left panel shows the \ac{PP} plot obtained using all 5000 simulated signals with (blue) and without (orange) the $\beta$ likelihood correction along with the expected 1-, 2-, and 3-$\sigma$ intervals.
We see that when including $\beta$, the credible interval \ac{CDF} for the luminosity distance deviates from the expected uniform distribution.
For both curves, we compute a $p$-value comparing the \ac{CDF} obtained for our simulated set of \acp{BBH} with the expected uniform distribution and verify that when excluding $\beta$, the recovery is unbiased ($p \geq 0.05$).
We note the blue curve in the left panel intersects zero on the vertical axis at $\approx 0.5$ on the horizontal axis; this is because the bias in~\ref{eq:biased-likelihood} does not impact the location of the peak of the likelihood, only the width.

Previous consistency checks of the performance of multiple inference libraries have failed to identify this bias.
To assess how many events are needed for a \ac{PP} test to identify the bias in this case, we calculate the $p$-value for a varying number of events in the right panel.
We see that hundreds of events are needed for the $p$-value to systematically fall to small values, which is more than the number of simulations that are typically analyzed when testing inference software due to computational limitations.

\begin{figure*}
\begin{center}
\includegraphics[width=0.49\linewidth]{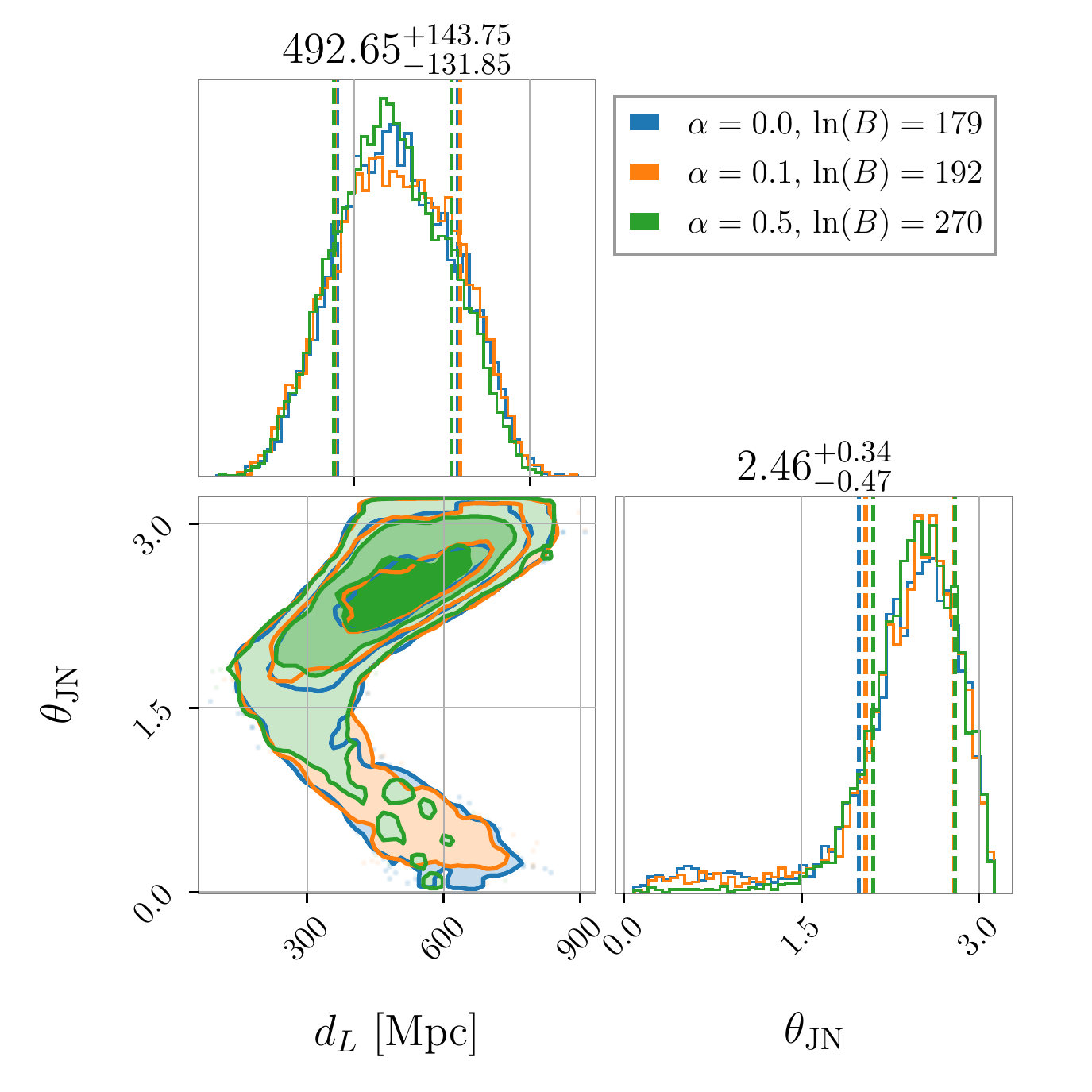}
\includegraphics[width=0.49\linewidth]{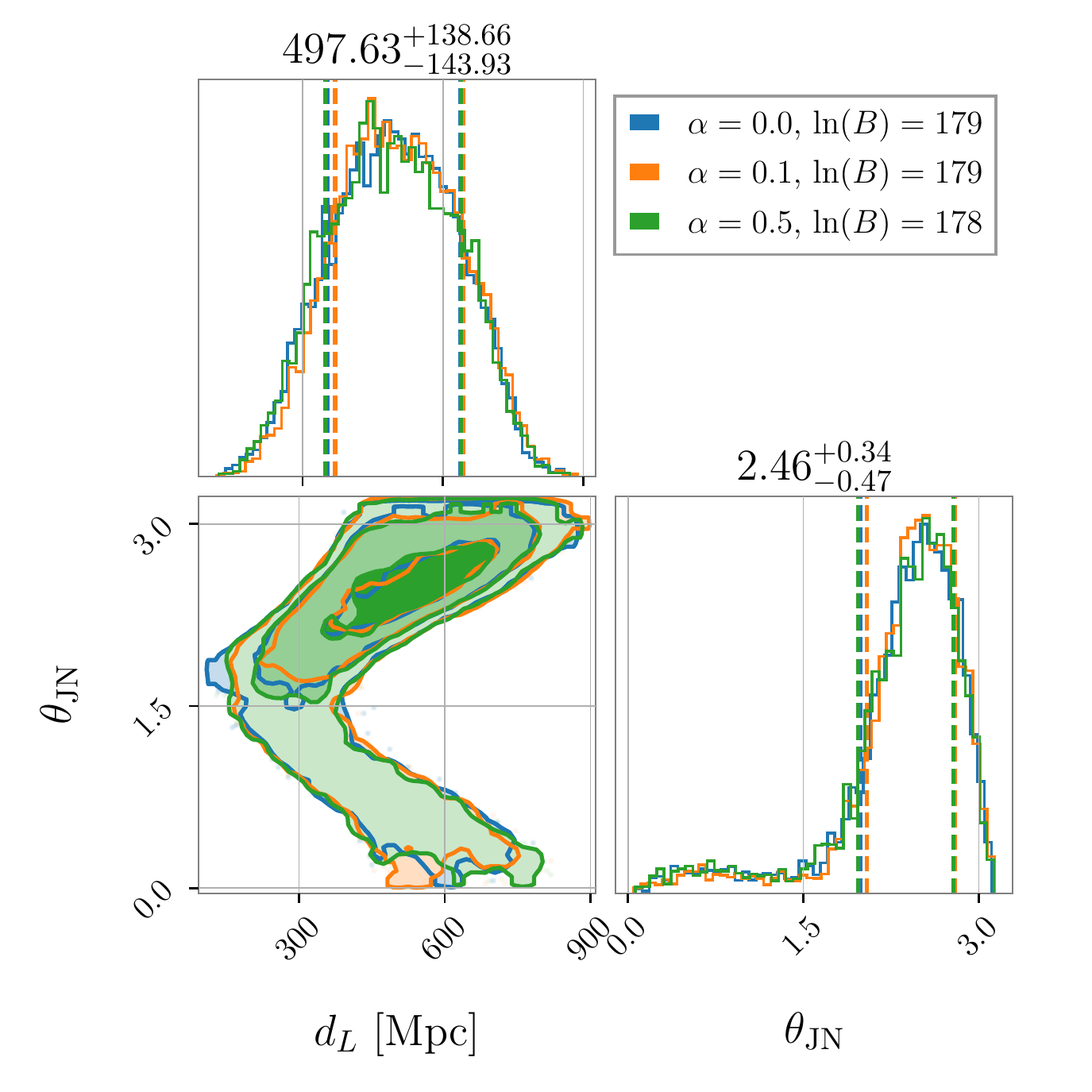}
\caption{
The one- and two-dimensional posterior probability distributions for luminosity distance and binary inclination angle obtained when analyzing GW150914 under a range of choices for the window function and likelihood.
The blue, orange, and green results are obtained using a Tukey window with $\alpha=0, 0.1, 0.5$, respectively.
These correspond to a rectangular window, the window function used during the original LVC analysis of GW150914, and the most extreme window used during \ac{O4}.
On the left, we use the previous definition of the likelihood (\ref{eq:biased-likelihood}) while on the right, we remove the factor that accounts for the window power loss $\beta$.
The legend shows the natural log Bayes factor comparing the signal vs Gaussian noise hypotheses.
On the right, we find that all the results agree well, even for the rectangular window.
On the left, we find that the log Bayes factor is higher by approximately $1 / \beta$ and the posterior is more strongly peaked when $\alpha$ is non zero.
}\label{fig:gw150914}
\end{center}
\end{figure*}

The second test we perform is an analysis of GW150914~\cite{LIGOScientific:2016aoc} using a range of values of the Tukey $\alpha$ with and without $\beta$ in the likelihood.
For computational efficiency, we ignore the impact of the spins of the two black holes and use the waveform {\sc IMRPhenomD}~\cite{Husa:2015iqa, Khan:2015jqa}, which neglects the impact of higher-order emission modes in the signal.
We perform the analysis using the likelihood defined in {\sc Bilby}~\cite{Ashton:2018jfp, Romero-Shaw:2020owr} and compute the Bayesian evidence and sample the posterior using the {\sc nestle} sampler~\cite{nestle}.
We analyze $4~\mathrm{s}$ of data taken from the gravitational-wave open science center~\cite{LIGOScientific:2019lzm} centred on the trigger time from the same source and use \ac{PSD} estimates from the LIGO-Virgo Collaboration's GWTC-1 data release produced using {\sc BayesWave}~\cite{Cornish:2014kda, ligo_scientific_collaboration_and_virgo_2021_5117703}.
We use a prior that is uniform in detector-frame component masses with limits in chirp mass and mass ratio ($25 < {\cal M}/M_{\odot} < 35$, $0.125 < q < 1$).
We set a prior on the peak time in the LIGO Hanford observatory with a width $0.1~\mathrm{s}$ on either side of the trigger time.
The other extrinsic parameters follow the expected geometric distributions with distance bounds at $[50, 2000]$ Mpc.

In Figure~\ref{fig:gw150914}, we show the one- and two-dimensional marginal posterior distributions for luminosity distance and the inclination of the orbital plane from the line of sight to the binary, along with the natural log Bayes factor comparing the signal hypothesis to Gaussian noise.
On the left, we use \ref{eq:biased-likelihood} and on the right, we remove the $\beta$ factor.
On the left, we notice two effects that we would expect if we are artificially inflating the likelihood.
The posterior distribution becomes systematically narrower when increasing $\alpha$ as evidenced by the shrinking $90\%$ credible regions, and the log Bayes factor increases by a factor $\approx 1 / \beta$.
Without the $\beta$ factor, there is no statistically significant difference in the inferred posterior distributions or log Bayes factors.

A perhaps surprising result here is that the analysis that uses a rectangular window produces a posterior and evidence estimate that are consistent with the other window choices without the $\beta$ correction.
It is widely held in the gravitational-wave data analysis community that correct results cannot be obtained when using rectangular windows~\cite{LIGOScientific:2019hgc} without performing the analysis in the time-domain~\cite{Isi:2021iql} (although see~\cite{Kou:2025bhk}).
This is due to significant spectral leakage when using a rectangular window (see Figure~\ref{fig:windows}).
However, as described above, a more relevant question is whether the data being compared with the signal model are correctly whitened.
Provided the signal model does not come close to the boundaries of the segment, the poorly whitened data near the edge of the segment does not impact the likelihood ratio.
While we find that using a rectangular window produces sensible results, we caution that avoiding interactions between the segment boundaries and the signal model can be challenging and the noise statistics will not follow the expected distribution (e.g., the whitened frequency-domain residuals will not follow a unit normal distribution).

\begin{figure*}
\begin{center}
\includegraphics[width=0.49\linewidth]{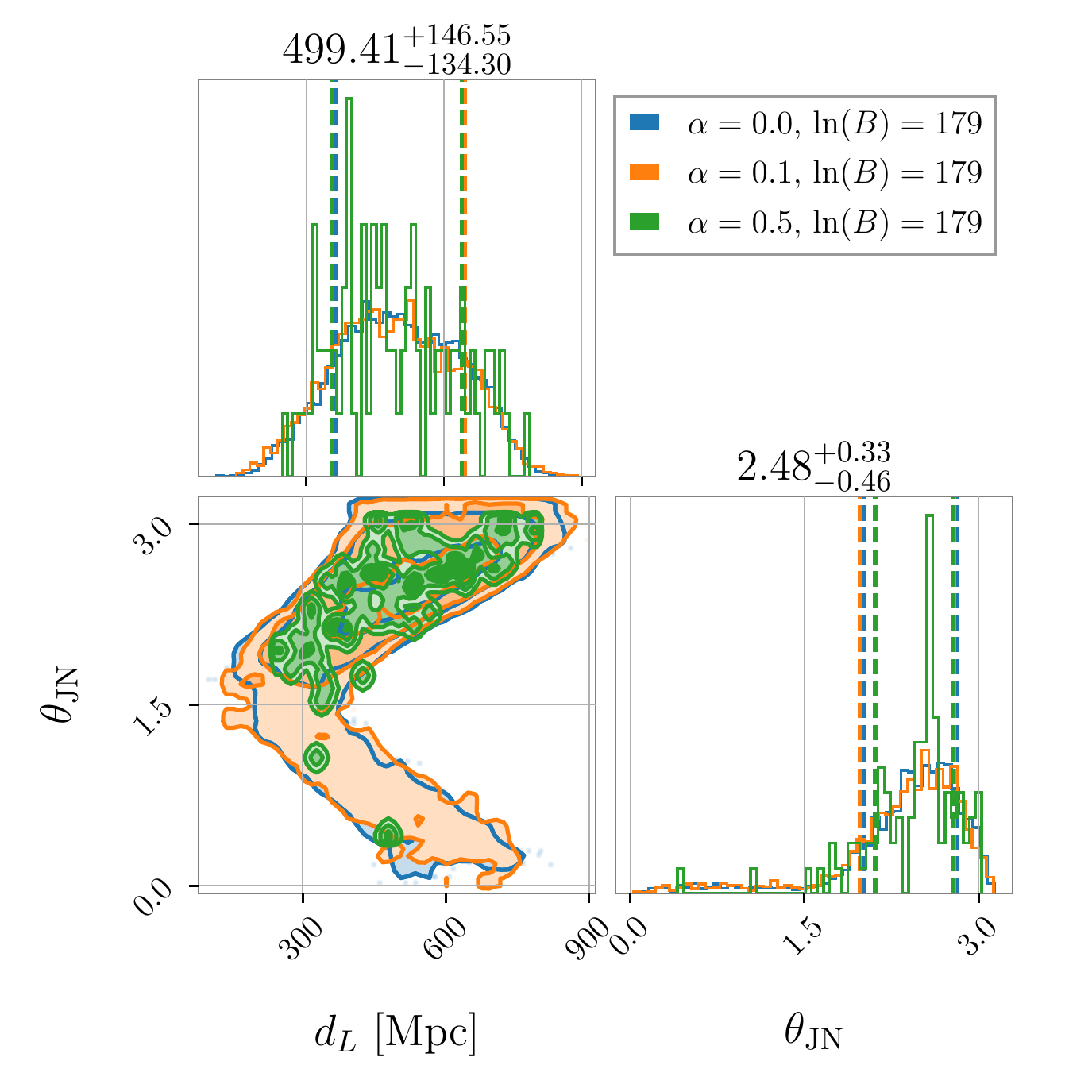}
\includegraphics[width=0.49\linewidth]{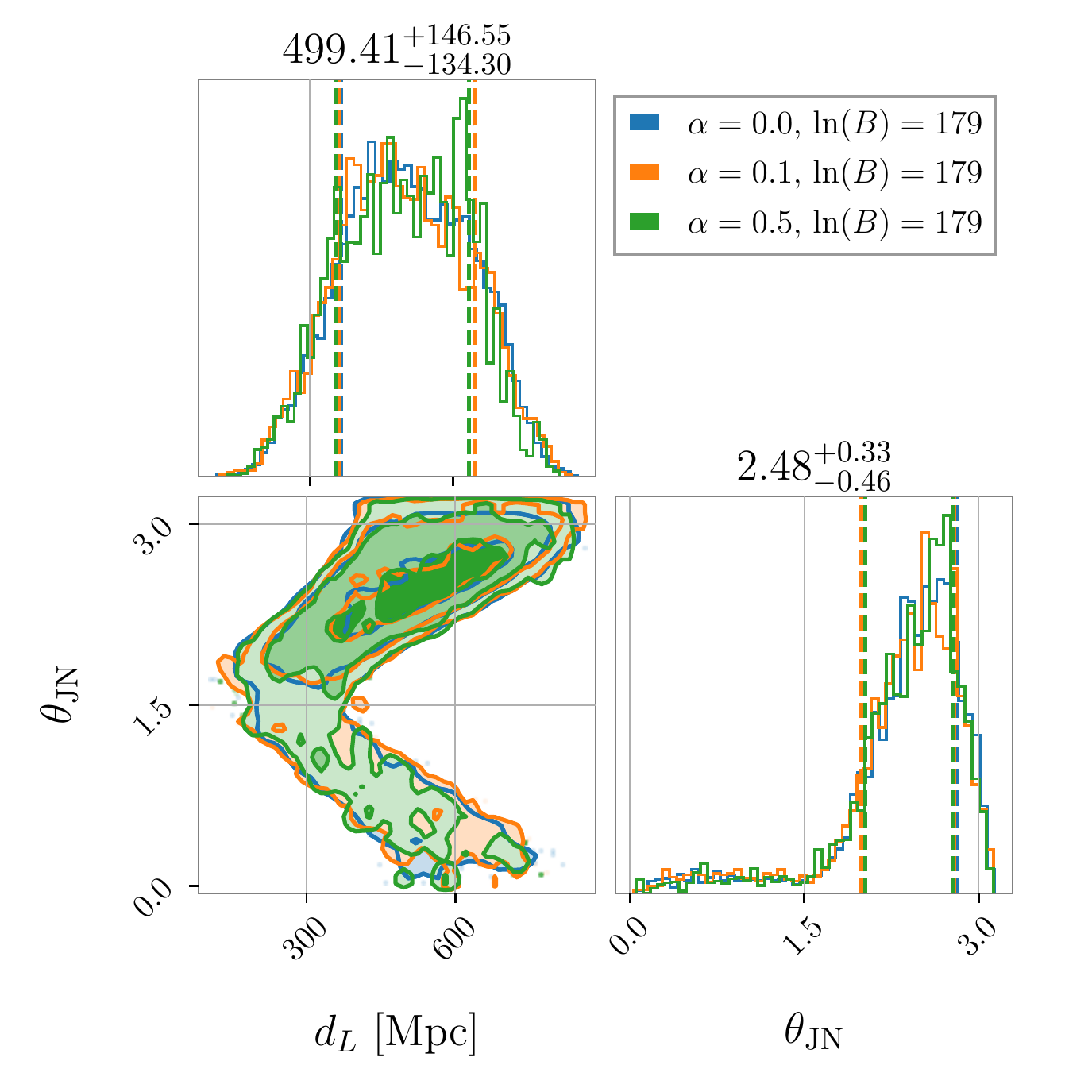}
\caption{
The one- and two-dimensional posterior probability distributions for luminosity distance and binary inclination angle obtained when analyzing GW150914 under a range of choices for the window function after correcting the biased results obtained using~\ref{eq:biased-likelihood} using rejection sampling.
The blue, orange, and green results are obtained using a Tukey window with $\alpha=0, 0.1, 0.5$, respectively.
These correspond to a rectangular window, the window function used during the original LVC analysis of GW150914, and the most extreme window used during \ac{O4}.
The legend shows the natural log Bayes factor comparing the signal vs Gaussian noise hypotheses.
On the left-hand side, we show the posterior samples obtained by rejection sampling the initial result, while on the right, we use repeated sampling with replacement in order to increase the effective sample size.
We find that using importance sampling, we correctly estimate the log Bayes factor in all cases.
On the right, we find that all the results agree well, although for $\alpha=0.5$ there is more variation in the posteriors.
However, when employing rejection sampling, the efficiency is extremely low for $\alpha=0.5$.
}\label{fig:gw150914-reweighted}
\end{center}
\end{figure*}

\subsection{Correcting results using importance sampling}

A common approach to modifying existing results in gravitational-wave astronomy is importance sampling~\cite[e.g.,][]{Payne:2019wmy}.
In our case, the importance sampling weights are straightforward to evaluate,
\begin{eqnarray}
    w_i &= \frac{{\cal L}(d | \theta_i)}{{\cal L}_{\beta}(d | \theta_i)} = {\cal L}_{\beta}(d | \theta_i)^{-\frac{5\alpha}{8}}.
\end{eqnarray}
We can use these weights to estimate correct Bayes factors $B = \langle w_i \rangle B_{\beta}$ and resample the posterior distribution according to the weights to estimate the corrected posterior.
For more details of the different resampling techniques we use, see~\ref{app:duplicate}.

In Figure~\ref{fig:gw150914-reweighted}, we show the posterior probability distributions obtained by importance sampling the results in the left panel of Figure~\ref{fig:gw150914} using two methods.
On the left, we use rejection sampling to obtain independent samples from the correct distribution and on the right, we sample with replacement until we have 80,000 samples.
For $\alpha=0.5$ only $0.7\%$ of samples are retained after rejection sampling, leading to a poorly sampled approximation of the posterior. 
Repeated sampling somewhat mitigates this, but the posterior still has larger uncertainty than the blue and orange results.
For $\alpha=0.1$, the rejection sampling efficiency is $\approx 40\%$.
The poor sampling efficiency in both cases is expected as importance sampling generally performs best when the target distribution is narrower than the initial distribution; however, we are in the opposite situation.
For all window functions, the log Bayes factor is correctly estimated using importance sampling.
Thus, we find that reweighting is a viable means to correct previous results obtained with the $\beta$ correction in the likelihood for all but the shortest signals analyzed with values of $\alpha \gtrsim 0.5.$

\section{Alternative solutions}
\label{sec:solutions}
Removing the $\beta$ factor from~\ref{eq:biased-likelihood} is only an approximate solution, and more complex methods are required to maintain unbiased inference.
Multiple solutions have been discussed previously in the literature, including modeling the full noise covariance matrix~\cite{Talbot:2021igi} or performing the analysis in the time-domain~\cite{Isi:2021iql}.
However, since these approaches require using the covariance matrix, they come at increased computational cost.

Another possible solution is to perform a multi-stage data conditioning routine that ignores the edges of the segment and ensures that the final windowing is applied to white noise.
To evaluate the likelihood using a section of data with duration $T$, the following steps can be taken:
\begin{enumerate}
    \item Apply a rectangular window to a section of data with duration $T + 2\Delta$ with $\Delta$ of extra data on each end and apply a discrete Fourier transform.
    \item Whiten the data in the frequency domain and inverse Fourier transform.
    \item Crop the whitened data to duration $T$ and apply a final Fourier transform.
\end{enumerate}
Signal models can then either be generated in the frequency domain with frequency spacing $1 / T$ and whitened before subtracting from the whitened and cropped frequency-domain data or generated with the full duration $T + 2 \Delta$ and processed using the same procedure as the data.
The latter has a larger computational cost ($O(N\log N)$ vs $O(N)$) but is more accurate for cases where the signal extends beyond the edge of the cropped segment.

Using the notation from Section~\ref{sec:formalism}, the likelihood can be written
\begin{eqnarray}
    \ln {\cal L}(d | \theta)
    &= \sum_{i \in \mathcal{S}, k \in \mathcal{K}, j, m, n} h_{i}(\theta) \mathcal W_{ij} P_{jk} P_{km} \mathcal W_{mn} D_{w,no} d_{o} .
\label{eq:correction_term_cropped}
\end{eqnarray}
Here $P_{jk}$ is a projection matrix from the full segment to the cropped segment and $k \in \mathcal{K}$ runs over the cropped $T$ duration segment, $\mathcal{S}$ is as before, and the remaining indices run over the entire $T + 2 \Delta$ duration segment.
The problematic term for this method is
\begin{eqnarray}
    \delta \ln {\cal L}(\theta)
    &= \sum_{i \in \mathcal{S}, k \in \mathcal{K}, j, m, n} h_{i}(\theta) \mathcal W_{ij} P_{jk} P_{km} \mathcal W_{mn} (\delta_{no}- D_{w,no})d_{o} .
\label{eq:correction_term_cropped}
\end{eqnarray}
The value of $\Delta$ should be chosen such that $\delta \ln {\cal L}$ in~\ref{eq:correction_term_cropped} is sufficiently small, so calculation of this quantity is a useful sanity check that can be performed either using prior samples or in post-processing using the posterior samples.

\begin{figure*}
\begin{center}
\includegraphics[width=0.49\linewidth]{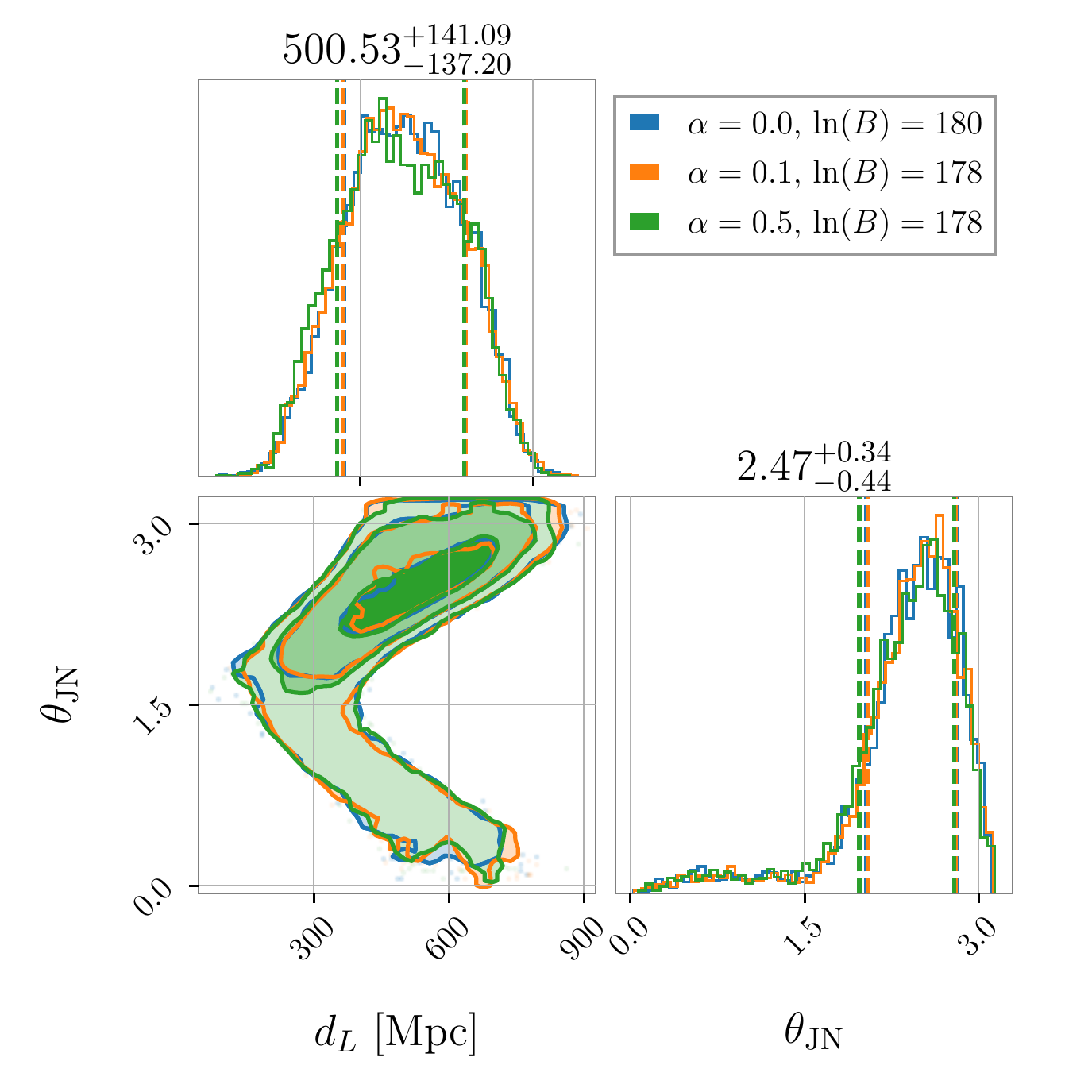}
\caption{
The one- and two-dimensional posterior probability distributions for luminosity distance and binary orientation obtained when analyzing GW150914 under a range of choices for the window function and our cropped likelihood.
The blue, orange, and red results are obtained using a Tukey window with $\alpha=0, 0.1, 0.5$, respectively.
These correspond to a rectangular window, the window function used during the original LVC analysis of GW150914, and the most extreme window used during the fourth observing run of the LIGO-Virgo-KAGRA detector network.
The legend shows the natural log Bayes factor comparing the signal vs Gaussian noise hypotheses.
We find that for all choices of window function, the results are consistent.
}\label{fig:gw150914-cropped}
\end{center}
\end{figure*}

As a demonstration of this method, we repeat our analysis of GW150914 using this cropped likelihood with $\Delta = 2~\mathrm{s}$.
The posterior distributions obtained using this likelihood with the same three choices of window function as in Figure~\ref{fig:gw150914} are shown in Figure~\ref{fig:gw150914-cropped}.
The posterior distribution and log Bayes factor obtained in each case is entirely consistent with that obtained with the standard likelihood and data conditioning procedure without the $\beta$ factor.

In Table~\ref{tab:ln-evidence}, we show the natural log evidence for the signal hypothesis for each of our twelve analyses.
For our cropped likelihood, the three log evidence values are similar; however, for the analyses that use the full stretch of whitened data, the choice of window has a significant impact on the recovered log evidence.
For a rectangular window, the absolute value of the log evidence is too large due to the non-periodicity of the analysis segment.
Using a Tukey window with a longer roll-off leads to a smaller magnitude for the log evidence, as some of the analyzed data has been suppressed.
When applying the $\beta$ correction, the log evidence values are approximately scaled by $\beta$, and after importance sampling, we recover the same log evidence values as when sampling without that term.
For perfectly whitened data, these values should follow a $\chi^{2}$ distribution with $N=3936$ degrees of freedom (up to a factor related to the prior volume).
We find that for the cropped data, the log evidences are slightly less than this, which we attribute to the PSD over-whitening the data at low frequencies and around the large lines at $\approx 500$ Hz.

As a final test of this new method compared to the Whittle likelihood, we calculate $\delta \ln {\cal L}$ from~\ref{eq:correction_term} for our analyses using the Whittle likelihood with each of our window functions and the cropped likelihood.
We take $64 s$ of data centred on the merger time and calculate $\delta \ln {\cal L}$ for the posterior samples from one of our unbiased analyses using data handling methods that match each of our unbiased likelihoods.
We find $\sigma_{\delta} \lesssim 0.1$ depending on $\alpha$ using the standard Whittle likelihood while for our cropped likelihood, we find $\sigma_{\delta} \lesssim 0.01$.
Since GW150914 has \ac{SNR} $\rho \sim 25$ we expected that we will not see a significant bias when $\rho \lesssim 250$ using the corrected Whittle likelihood.
In contrast, the bias for the cropped likelihood will be sufficient for ${\rm SNR} \lesssim 2500$.
We note that neglecting higher-order modes in this calculation may make these limits overly optimistic.
We leave further testing of this cropped likelihood for deployment at scale to future work.

\begin{table}
\caption{
    The natural log evidence for the signal hypothesis for each analysis of GW150914 presented in this work.
    We see that the cropped likelihood yields similar log evidence values for each choice of window function.
    In contrast, the choice of window has a very significant impact on the estimated evidence for both of the other likelihoods.
    When including $\beta$ the values are systematically higher than without $\beta$, as expected.
}
\label{tab:ln-evidence}
\begin{center}
\begin{tabular}{@{\extracolsep{\fill}}lccc}
\br
& $r = 0\,\mathrm{s}$ & $r = 0.2\,\mathrm{s}$ & $r = 1\,\mathrm{s}$ \\
\mr
with $\beta$       & $-51847$ & $-4648$ & $-3442$ \\
with $\beta$ reweighted & $-51847$ & $-4358$ & $-2372$ \\
without $\beta$    & $-51848$ & $-4358$ & $-2373$ \\
cropped            & $-3264$  & $-3248$ & $-3253$ \\
\br
\end{tabular}
\end{center}
\end{table}

\section{Conclusion}
\label{sec:conclusion}

In the decade since the first generation of \acp{GW}, the increasing sensitivity of \ac{GW} detectors and the quantity of signals observed have increased the precision requirements for Bayesian inference with these sources.
Several sources of bias have been discussed in the literature including the impact of imperfect detector calibration~\citep{Huang:2022rdg, Essick:2022vzl}, uncertainties in \ac{GW} signal models~\citep{Purrer:2019jcp}, and mismodeling of detector noise~\citep{Payne:2022spz, Hourihane:2025vxc, Chatziioannou:2019zvs, Biscoveanu:2020kat, Talbot:2020auc, Plunkett:2022zmx}.
In this paper, we have described a widespread misconception in data analysis for gravitational wave transients stemming from a modification of the likelihood to incorrectly account for power lost in the data due to windowing in the time domain and described a method to quantify the bias in our inference due to our choice of window function.

We demonstrate the impact of this error on gravitational-wave parameter estimation through a series of numerical experiments which imply that all previously-published individual-event parameter estimation produced by the \ac{LVK} has used an incorrectly over-constrained likelihood, $\mathcal{L}_{\beta}(d | \theta) = \mathcal{L}(d | \theta)^{1/\beta}$.
Instead, we advocate for the use of the Whittle likelihood without the $\beta$ correction as an immediate solution and show that this version of the likelihood\footnote{As of version 2.6.0 of \textsc{Bilby}~\cite{colm_talbot_2025_15918940} the $\beta$ factor has been removed.} leads to unbiased inference for \acp{SNR} $\lesssim O(100)$ in the case where only the signal-to-noise likelihood ratio is of interest.
However, this uncorrected form of the likelihood does not provide unbiased estimates of the signal evidence and absolute likelihood. 
We also present a more general method that better models the absolute likelihood of the analyzed data using a multi-stage data conditioning procedure at minimal additional computational cost that should be sufficiently accurate for \acp{SNR} $\lesssim O(1000)$.

\ack
\label{sec:acknowledgments}

This material is based upon work supported by NSF's LIGO Laboratory which is a major facility fully funded by the National Science Foundation. LIGO was constructed by the California Institute of Technology and Massachusetts Institute of Technology with funding from the National Science Foundation and operates under cooperative agreement PHY-0757058. 
C. T. is supported by the Eric and Wendy Schmidt AI in Science Postdoctoral Fellowship, a Schmidt Sciences program.
S. B.~is supported by NSF PHY-2513246 and by NASA through the NASA Hubble Fellowship grant HST-HF2-51524.001-A awarded by the Space Telescope Science Institute, which is operated by the Association of Universities for Research in Astronomy, Inc., for NASA, under contract NAS5-26555.
A. Z.~is supported by NSF Grant PHY-2308833.
T.B.~is supported by the research program of the Netherlands Organisation for Scientific Research (NWO).
C.H.~thanks the UKRI Future Leaders Fellowship for support through the grant MR/T01881X/1.
M.J.W.~acknowledges support from STFC grants ST/X002225/1, ST/Y004876/1 and the University of Portsmouth.
J. V.~acknowledges support from STFC grant ST/V005634/1.
The 1 thank Gregory Ashton for insightful discussions and comments on the manuscript.
The authors are grateful for computational resources provided by the Caltech LIGO Laboratory supported by NSF PHY-0757058 and PHY-0823459.
This paper carries LIGO document number LIGO-P2500488 and preprint number UT-WI-24-2025.
The {\sc Jupyter} notebooks used to produce the results in this paper are on GitHub at \href{https://github.com/ColmTalbot/windowing-in-parameter-estimation}{github.com/ColmTalbot/windowing-in-parameter-estimation}.
This work used the following software packages: {\sc numpy}~\cite{harris2020array}, {\sc scipy}~\cite{2020SciPy-NMeth}, {\sc matplotlib}~\cite{Hunter:2007}, {\sc corner}~\cite{corner}, {\sc lalsimulation}~\cite{lalsuite}, {\sc gwpy}~\cite{gwpy}, {\sc nestle}~\cite{nestle}, {\sc Bilby}~\cite{Ashton:2018jfp,Romero-Shaw:2020owr}.

\bibliographystyle{iopart-num}
\bibliography{citations.bib}

\appendix

\section{Comparison of importance sampling methods}\label{app:duplicate}

For a set of weighted samples with weights $w_{i}$ one can define the effective sample size
\begin{equation}
    {\rm ESS} = \frac{\left(\sum^{N}{w_i}\right)^2}{\sum^{N}{w_i^2}}
\end{equation}
with limits ${\rm ESS} = N$ for $N$ equally weighted samples and ${\rm ESS} = 1$ when only one sample is nonzero.
This effective sample size is a useful heuristic when estimating the reliability of Monte Carlo integrals computed, e.g., in our Bayes factor calculation, or visualizing posteriors using weighted samples.

However, it is often simpler not to carry weights around, so we resample according to the weights.
There are two common resampling methods, with or without duplication (also referred to as replacement).
The most common method of resampling without replacement is rejection sampling, wherein each weight is compared with a random draw from $r_i \sim U(0, w_{\max})$ and accepted if $w_i > r_i$.
This method leads to an average efficiency of
\begin{equation}
    \bar{\epsilon}_{{\rm rejection}} \approx \frac{\sum^{N}{w_i}}{N w_{\max}}.
\end{equation}
Since rejection sampling yields equally weighted, independent samples, the effective sample size after rejection sampling is $\bar{N}_{\mathrm{rejection}} = N \bar{\epsilon}_{\mathrm{rejection}}$.
Generically, we can see $N_{\mathrm{rejection}} \leq \mathrm{ESS}$.
To show this, we write
\begin{eqnarray}
    \frac{\rm ESS}{\bar{N}_{\rm rejection}} &= \frac{\left(\sum^{N}{w_i}\right)^2}{\sum^{N}{w_i^2}} \frac{w_{\max}}{\sum^{N}{w_i}} = \frac{\left(\sum^{N}{w_{\max} w_i}\right)}{\sum^{N}{w_i^2}} \geq 1.
\end{eqnarray}
In the last step, we use $w_{\max} \geq w_i$.

When sampling with replacement, we also have equally weighted samples; however, they are not independent.
We can calculate the effective sample size in this case by drawing $M$ samples and counting the number of unique samples $K$ and occurrence of each $k_{i}$.
This is equivalent to drawing $K$ samples with weights $k_{i}$ and so we have
\begin{eqnarray}
    N_{\mathrm{repeated}}
    = \frac{\left(\sum^{K}{k_i}\right)^2}{\sum^{K}{k_i^2}}
    = \frac{M^2}{\sum^{K}{k_i^2}}.
    = \left(\sum^{K}\left(\frac{k_i}{M}\right)^2\right)^{-1}
    .
\end{eqnarray}
We can see that this is a discretized approximation to $\mathrm{ESS}$ and so $N_{\mathrm{repeated}} \rightarrow \mathrm{ESS}$ as $M \rightarrow \infty$.

\end{document}